\documentclass[a4paper,jpcm]{iopart}
\usepackage{graphicx}
\usepackage{subfigure}
\begin{document}
\title[(Ga$_{1-x}$In$_x$)$_2$O$_3$ over the whole $x$ range]{Properties of 
(Ga$_{1-x}$In$_x$)$_2$O$_3$ over the whole $x$ range
}

\author{M B Maccioni, F Ricci and V Fiorentini}

\address{CNR-IOM and Dept. of Physics, University of Cagliari, Cittadella Universitaria, 09042 Monserrato (CA), Italy }

\ead{vincenzo.fiorentini@dsf.unica.it}

\begin{abstract}
Using  density-functional ab initio  theoretical techniques, we study  (Ga$_{1-x}$In$_x$)$_2$O$_3$  in both its equilibrium structures (monoclinic $\beta$ and bixbyite) and over the whole range of composition. We establish that the alloy exhibits  a large and temperature-independent miscibility gap. On the low-$x$ side, the favored phase is isostructural with $\beta$-Ga$_2$O$_3$; on the high-$x$ side, it is isostructural with bixbyite In$_2$O$_3$. The miscibility gap opens between approximately 15\% and 55\% In content for the bixbyite alloy grown epitaxially on In$_2$O$_3$, and 15\% and 85\% In content for the free-standing bixbyite alloy. The gap, volume  and band offsets to the parent compound also exhibit anomalies as function of $x$. Specifically, the offsets in epitaxial conditions are predominantly type-B staggered, but  have  opposite signs in the two end-of-range phases.
\end{abstract}

\section{Introduction}
The group-III sesquioxides  Ga$_2$O$_3$ and In$_2$O$_3$ are currently popular in basic materials science and technology for being, respectively, deep-UV large-breakdown  and  near-UV transparent-conducting materials. 
A natural development that can be envisaged is the  growth of a solid solution (Ga$_{1-x}$In$_x$)$_2$O$_3$, typically (but not necessarily) epitaxially on the parent compounds. This would enable one to combine and tune the functionalities of the two parent compounds, and to export the band-engineering and nanostructuration concepts well known in other semiconductor systems (such as arsenides and nitrides) to much  higher absorption energies and breakdown voltages. 

Progress in this directions requires a knowledge of the miscibility, as well as the behavior of relevant properties (gap, specific volume, band offsets, etc.),  of a solid-solution substitutional alloy composed, in a given proportion, of the two parent materials. In this paper we report the modeling of  (Ga$_{1-x}$In$_x$)$_2$O$_3$ over the entire range of $x$ by (mostly)  ab initio density-functional-theory techniques. Previous results \cite{noi1,noi2} for the low-$x$ end of the composition range are integrated in a comprehensive picture of the miscibility and attendant properties. The main result is that the alloy will phase-separate in a large composition range. On the low $x$ side, the favored phase is isostructural with $\beta$-Ga$_2$O$_3$; at large $x$, it has a bixbyite structure. We also find that as function of $x$ the gap, volume, and band offsets to the parent compound exhibit discontinuities typical of a first-order phase transition as function of  $x$. 

\section{Methods and technical issues}
Geometry and volume optimizations as well as electronic structure calculations are performed using density-functional theory (DFT) in the generalized gradient approximation (GGA), and the Projector Augmented-Wave (PAW) method as implemented in the VASP code \cite{vasp}.

For mixtures of  insulators, it is almost invariably the case that the structure of the mixture is that (or one of those) of the parent compounds. (This certainly does not generally apply to metal alloys.)  The stable phase for Ga$_2$O$_3$ is monoclinic $\beta$,  for In$_2$O$_3$ it is bixbyite. These are discussed at length in previous works (\cite{noi1} and \cite{sesqui1}, respectively). While there are no competing structures for In$_2$O$_3$, for Ga$_2$O$_3$ the $\beta$ phase can be transformed under pressure into the $\alpha$ phase; thus we checked these latter phases, and the bixbyite structures. We find that the $\beta$ phase is lowest in energy; the bixbyite is  80 meV/formula unit higher, and the $\alpha$ is 250 meV/formula unit higher. We therefore neglect the latter and  consider the $\beta$  and the bixbyite structures.

For the $\beta$ structure we use, as in \cite{noi1},  supercells containing 1$\times$2$\times$2 unit cells (80 atoms) and a 4$\times$4$\times$2 k-point grid. For the bixbyite structure, we use the cubic primitive cell (80 atoms), and a 4$\times$4$\times$4 k-point grid. The plane wave cutoff is 300 eV, exceeding by 10\% or more for all PAW sets the maximum suggested value. The structure of the $\beta$ phase,  reported in \cite{noi1}, is in good agreement with experiment; for bixbyite the lattice constant is 10.29 \AA, or 1.5\% larger than experiment as usual for GGA calculations.

\begin{figure}[ht]
\centering
\qquad\includegraphics[width=9cm]{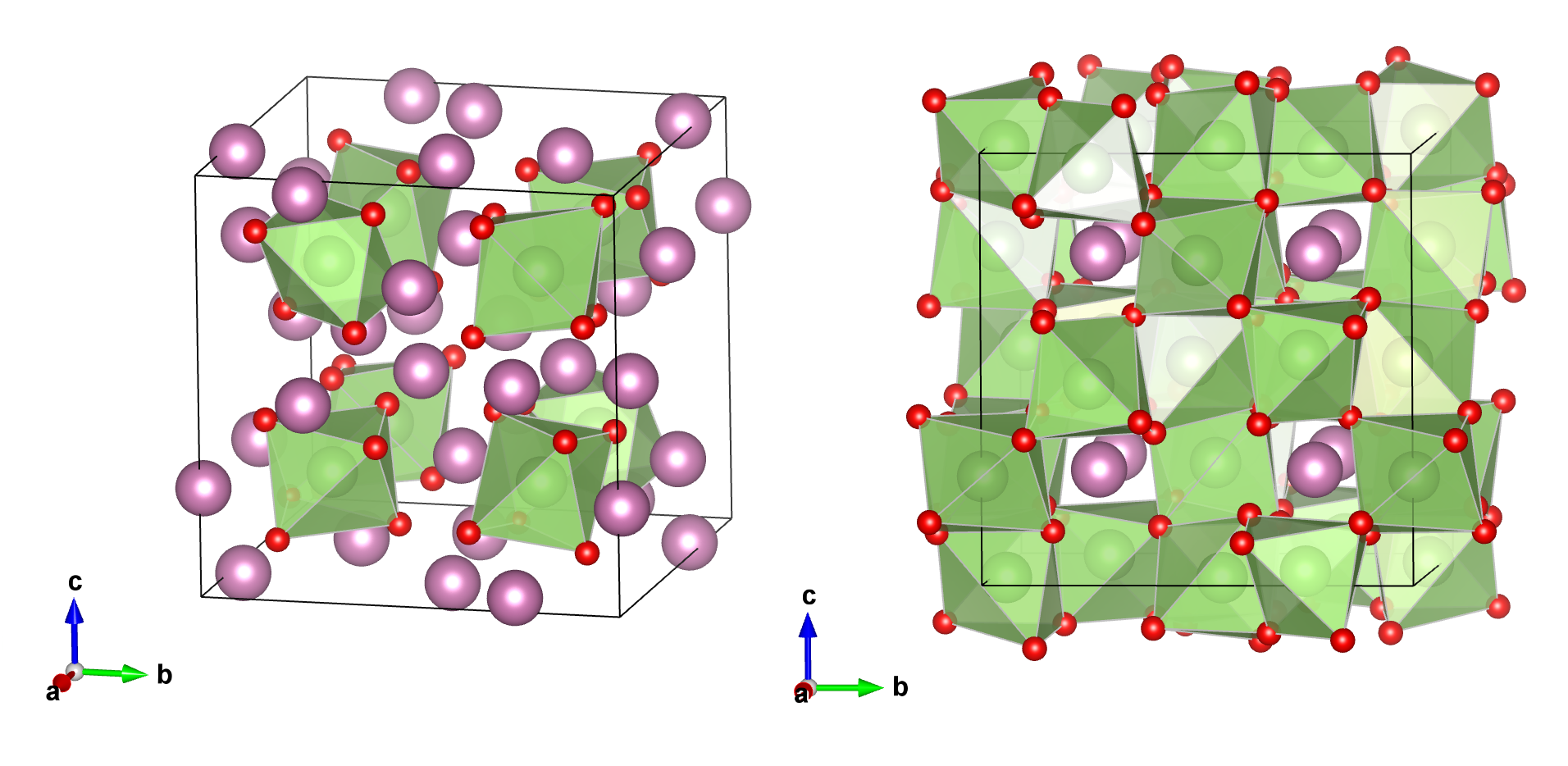}\qquad
\caption{\label{fig1} The bixbyite  structure (group T$_h$) has six-fold coordinated cations (large spheres) occupying 8$b$ high-symmetry and 24$d$ Wyckoff sites. In the left panel we highlight the 8$b$ cation sites by surrounding them with the local octahedra with oxygen (small spheres) at their vertexes; in the right panel we  do the same with the 24$d$ lower-symmetry cation sites.}
\end{figure}

We simulate the compositional variation explicitly mixing  In and Ga cations, as dictated by the mole fraction $x$ of In. For low $x$, we consider the  monoclinic $\beta$-Ga$_2$O$_3$ phase doped with In. This alloy is free-standing, i.e. its energy is calculated at zero stress; we have checked that epitaxial constraints on this phase do not change any of our conclusions, so we neglect them here for clarity. As found  previously \cite{noi2}, this phase is only relevant up to about $x$=0.25. We then  study the bixbyite phase over the whole range of $x$; this is obtained  naturally substituting Ga for In in In$_2$O$_3$, which is indeed a bixbyite as many other sesquioxides \cite{sesqui1,sesqui2}. Specifically, we study the bixbyite alloy in two settings: free-standing and In$_2$O$_3$-epitaxial. In the latter, the in-plane lattice parameters are fixed to that of In$_2$O$_3$ and the vertical lattice parameter and all internal coordinates are optimized.

Both the $\beta$ and bixbyite supercells contains 80 atoms, i.e. 32 cations. The choice of configurations in the $\beta$ phase has been discussed in Ref.\cite{noi1}. For the bixbyite phase, we find that Ga substitution is slightly favored at the high-symmetry cation site  (see Fig.\ref{fig1}). We substitute  Ga for In on those sites first, and then on the lower symmetry ones. For each $x$, we sample a few configurations; as Ga's generally try avoid each other, the geometric constraints on the  configurations are stringent, so there are not many possible inequivalent configurations to begin with. Our discussion is based on the lowest energy found at each $x$; of course, these may well not be the absolute energy minima for that $x$. Also, we neglect the possible occurrence of higher-energy configurations  in small proportions  at finite temperature.

\section{Phase separation}
\label{phsep}
To address the occurrence of phase separation, we calculate the specific (i.e. referred to one cation) Helmholtz free energy (the enthalpy  vanishes because the pressure is alway  zero to numerical accuracy)  of the mixture as a function of $x$. The internal energy is calculated directly from first principles using the supercells described above.  The entropy is modeled as the sum of mixing and vibrational terms. The mixing entropy has the standard form
\begin{equation}
S_m (x) = -x\, \log{x} - (1-x)\, \log{(1-x)}.
\end{equation}
Since growth  temperatures are comparable to or higher than Debye temperatures ($\Theta_{\rm In_2O_3}$=420 K, $\Theta_{\rm Ga_2O_3}$=870 K), the vibrational entropy per cation can  be approximated as that of a single-frequency oscillator at the Debye frequency. Thus 
\begin{equation}
S_v (x) = 3\,\left[(1+n)\, \log{(1+n)}-n\, \log{n}\right], 
\end{equation}
where $n$ is the Planck-Bose distribution
\begin{equation}
n (T,x) = 1/(e^{\Theta_m(x)/T}-1),
\end{equation}
and the mixture's  Debye temperature $\Theta_m$($x$) is assumed to be an interpolation \begin{equation}
\Theta_m (x)  = (1-x)\, \Theta_{\rm Ga_2O_3} + x\, \Theta_{\rm In_2O_3}
\end{equation}
 of that of the parent compounds.

We now recall that phase separation in a mixture occurs when the  specific free energy is a negative-curvature function of an extensive parameter such as  $x$. The values, say, $x_1$$<$$x_2$, at which the curvature becomes negative and goes back to positive, respectively, delimit the phase separation region; in general these bounds depend on temperature T, and the $x$ range they identify is the miscibility gap. If the negative curvature region vanishes as T increases, i.e. $x_1$ and $x_2$ get to coincide, there is complete miscibility. Our results, as mentioned earlier and discussed below, suggest a large miscibility gap  surviving up to above the melting temperatures of the parent compounds.
We discuss our results in terms of  the {\it mixing} specific free energy, i.e we subtract out the bulk free energy 
\begin{equation}
F_{\rm bulk} (x) = x\, F_{\rm In_2O_3} + (1-x)\, F_{\rm Ga_2O_3},
\end{equation}
which interpolates the values for two equilibrium bulk phases (bixbyite and $\beta$, respectively).

\begin{figure}[ht]
\centering
\includegraphics[width=6cm]{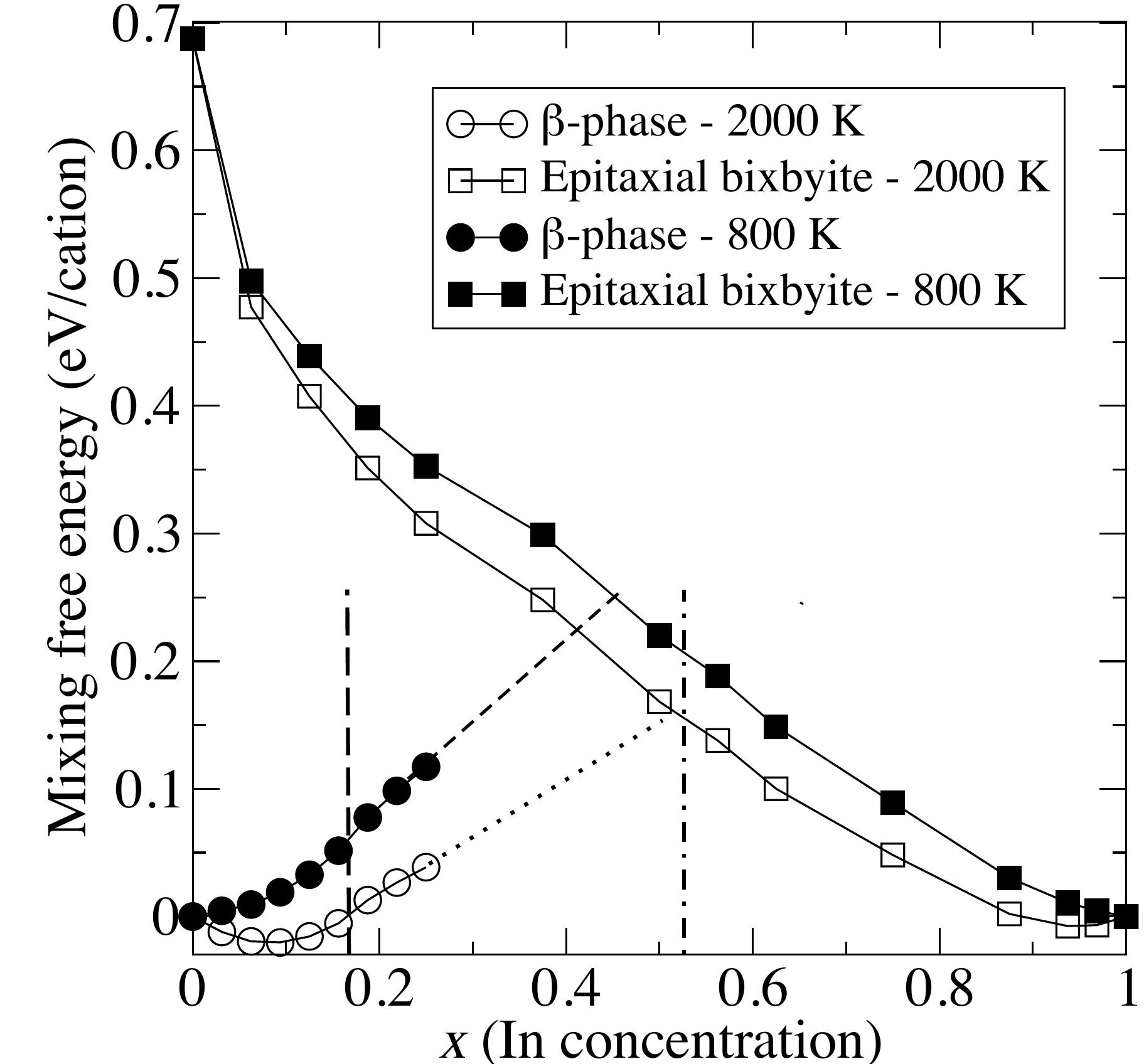}\qquad
\includegraphics[width=6cm]{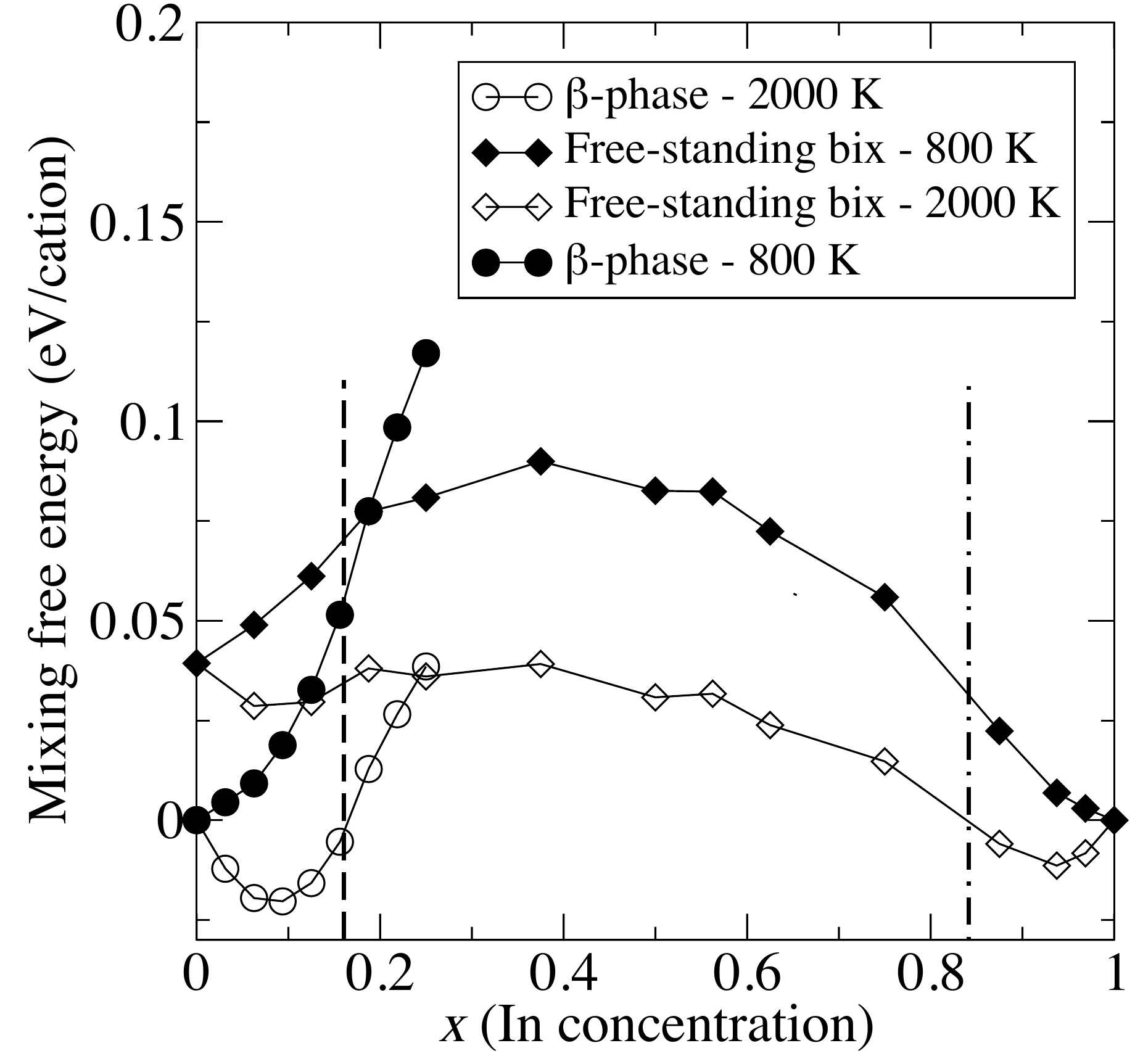}\qquad
\centering
\caption{\label{fig2}
 Mixing free energy vs $x$  at 810 K and 2000 K for $\beta$-phase vs epitaxial bixbyite (left panel), and $\beta$-phase vs free-standing bixbyite (right panel). The phase separation region extends between the vertical dashed and dash-dotted lines.}
\end{figure}

Fig.\ref{fig2}, left panel, compares the mixing free energies of the free-standing $\beta$ phase (circles) with that of the epitaxial bixbyite phase (squares).
Fig.\ref{fig2}, right panel, compares the same quantities for the same $\beta$ phase (circles) with that of the 
 free-standing bixbyite phase (diamonds). The temperatures considered are 800 K, a typical growth temperature, and 2000 K, near the melting temperature of the parent compounds. The free energy is evidently upward-convex in a wide region of intermediate $x$, indicating that a phase separation occurs.  The borders of that region define the miscibility gap.

On the low-$x$ end, the $\beta$ phase prevails in all cases, and the change in curvature occurs (hence the phase separation region starts) at about $x$$\simeq$0.15. This confirms largely our previous estimate \cite{noi1} of 10\% maximum In solubility, and experiments \cite{baldini} giving similar results.
At high $x$, the end of the miscibility gap region is  estimated at $x$$\simeq$0.45$\div$0.55 for the epitaxial case (left panel), subject to large uncertainties in locating the free-energy downturn from the epitaxial phase. 
Therefore the miscibility gap is approximately $x$$\in$(0.15,0.55) for the epitaxial bixbyite and $\beta$ phase. 
Comparison with the growth and x-ray diffraction study by Zhang {\it et al.} \cite{zhang} suggest that this prediction is quite plausible, even accounting for their epitaxial conditions being different from those simulated. 

Most importantly, at 2000 K the borders of the phase separation region are about the same as at 800 K, i.e. the miscibility gap  hardly changes  (it actually may  widen slightly).  Since the melting temperatures of the parent compounds are around 2200 K, we conclude that in the practical range of T the miscibility gap between the epitaxial and $\beta$ phases is $x$$\in$(0.15,0.55) independent of  T. 

We now come to  the $\beta$ phase vs free-standing bixbyite competition (Fig.\ref{fig2}, right panel). A phase separation region exists here too,  involving the structure change to the $\beta$ phase at low $x$: the lower limit is again $x$=0.15. The free-standing bixbyite phase is  favored over the $\beta$ phase (as well as over the epitaxial) over the rest of the $x$ range,  from $x$=0.2 or so onward.  However, its own free energy is upward-convex for most of the range; this  indicates a phase separation between   Ga$_2$O$_3$ and In$_2$O$_3$ {\it within} the bixbyite phase; the change in curvature on the high $x$ side is approximately at $x$=0.8$\div$0.85. Therefore the overall miscibility gap is $x$$\in$(0.15,0.85) in the free-standing case. This is quite clearly the case at both 800 K and 2000 K. Thus, as in the epitaxial case, we conclude that in the practical range of T the miscibility gap for free-standing bixbyite is $x$$\in$(0.15,0.85) independently of  T. 
An experimental determination of the Ga$_2$O$_3$-In$_2$O$_3$ phase diagram \cite{galliaindia} suggests that indeed at the In end of the range the single-crystal stability region is quite marginal, being limited to $x$$>$0.9 or so.

\section{Gap and volume}

\begin{figure}[ht]
\centering
\includegraphics[width=6.3cm]{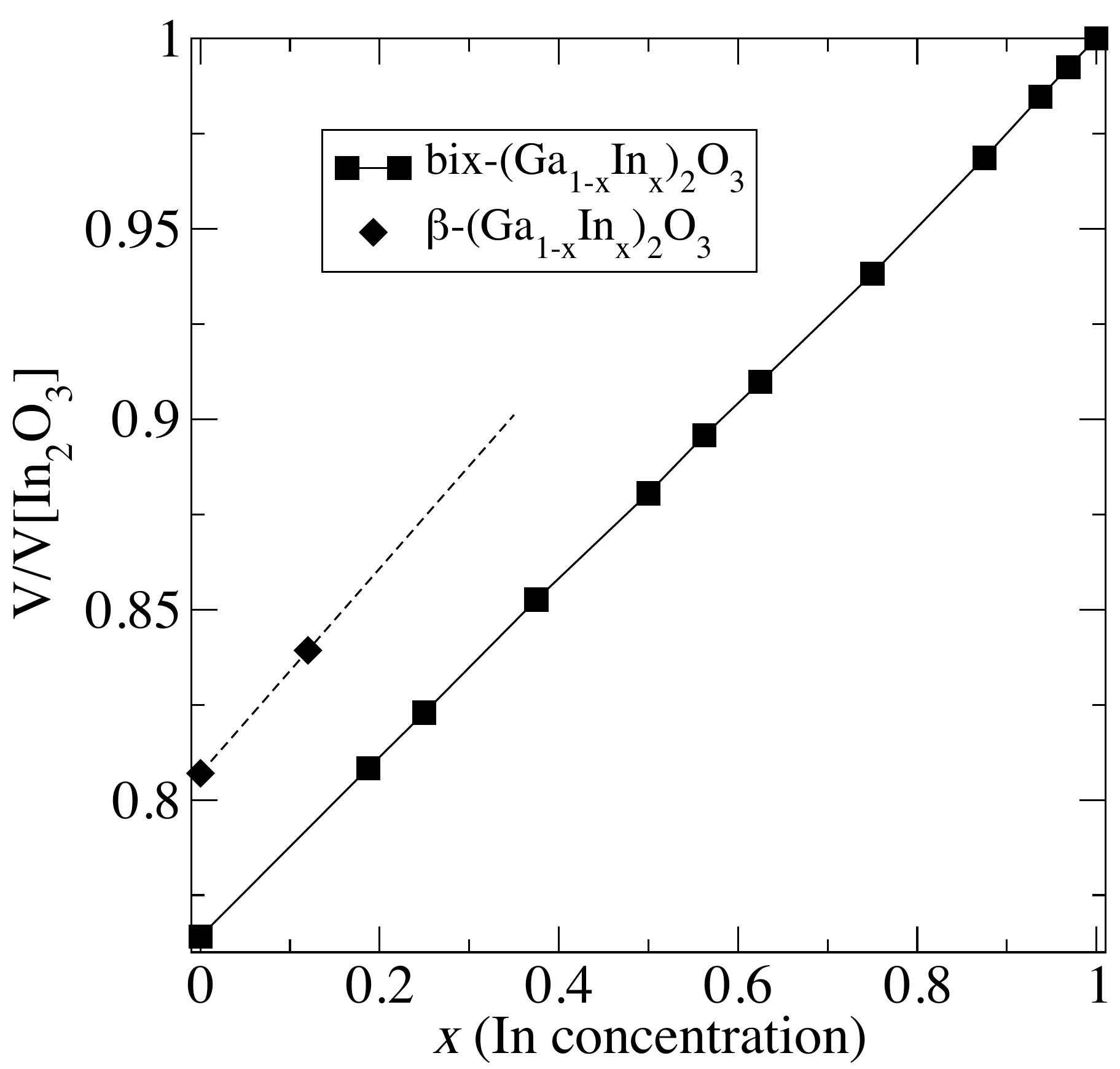}\qquad
\includegraphics[width=5.85cm]{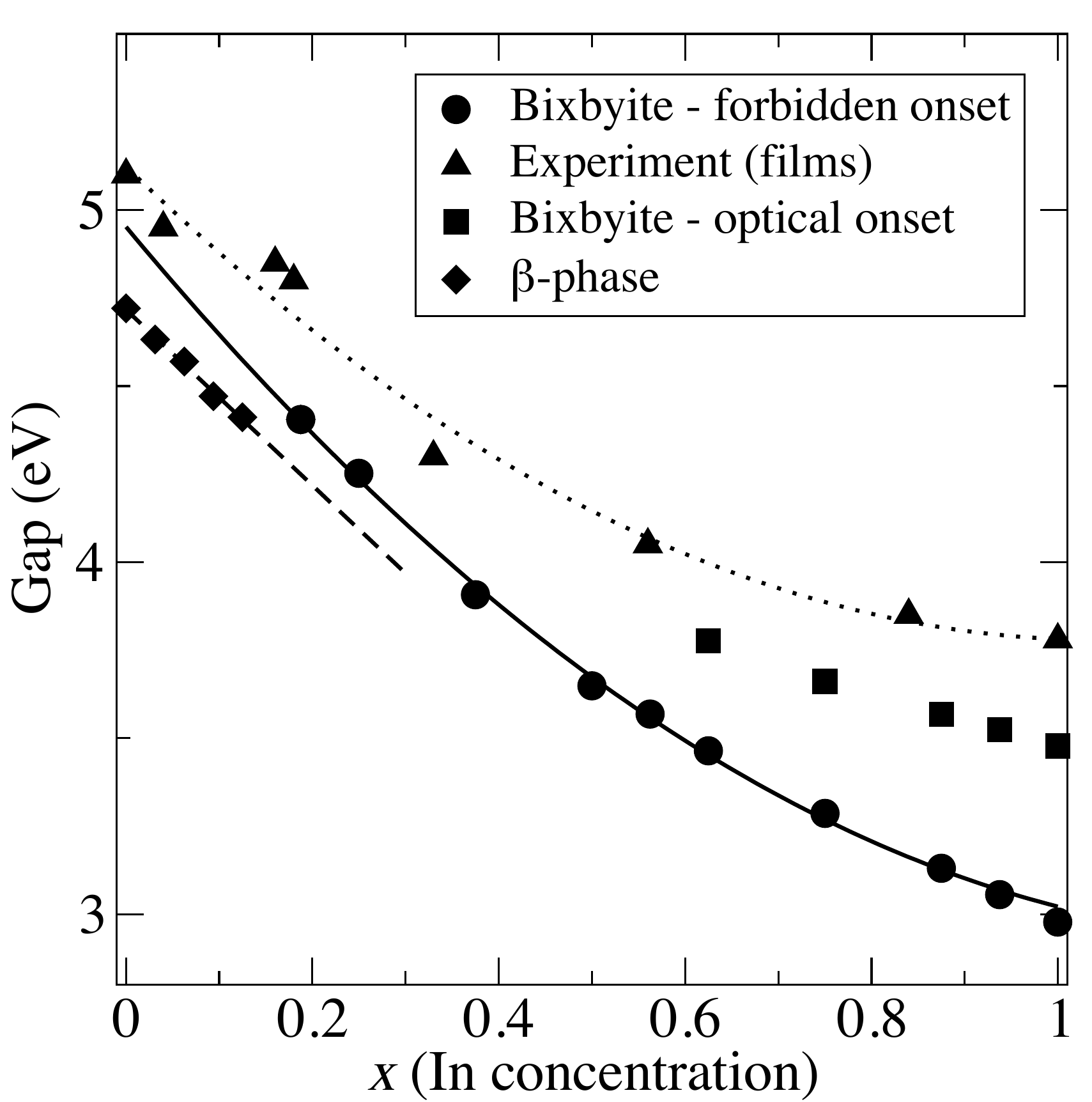}\qquad
\caption{\label{fig4} Left panel: volume vs mole fraction for the free-standing $\beta$ and bixbyite phases. Right panel: fundamental gap in the same phases and interpolations vs $x$ (quadratic for bixbyite; linear at low $x$ for $\beta$). A correction for the gap error has been introduced (see text).  The gap show a sizable bowing in bixbyite at large $x$.}
\end{figure}

Not unexpectedly, the properties of the alloy  exhibit anomalies  as function of concentration due to the changes in phase and structure. In Fig.\ref{fig4}, left panel, we show the calculated optimized volume in the two free-standing bixbyite and $\beta$ phases, showing a clear  volume discontinuity at any given $x$. The fundamental  gap, shown in Fig.\ref{fig4}, right panel, also exhibits analogous interesting features. The $\beta$ phase has a linear decrease in good agreement with pressure experiments \cite{noi2,zhang}. The bixbyite gap is also linear at low $x$, but picks up a significant bowing near $x$=1. To correct for the semilocal density-functional error, we supplement the GGA calculated gap with an ad hoc ``scissor"-like  correction 
\begin{equation}
\delta E_g(x)=2.5\,x+2.7\,(1-x)\ \rm eV,
\label{delta}
\end{equation}
 which brings the GGA gap to  the experimental value in Ga$_2$O$_3$ and In$_2$O$_3$ \cite{noi3,inogap} (incidentally, the correction reduces the bowing as obtained from GGA eigenvalues).  Since the lowest gap is dipole-forbidden, to compare with  the experimental optical onsets \cite{zhang}  we estimate  the position of the optical onset at  all $x$  as the GGA gap value (corrected by Eq.\ref{delta}) plus the  difference of optical onset and  minimum gap in In$_2$O$_3$  (0.55 eV). 
 
The   agreement is decent, but should  be taken with more than a grain of salt: on the experiment side, the data are for  films grown on sapphire, the In content is generally lower than the nominal one especially at intermediate $x$, and the gap in the $x$=0 and $x$=1 limits is larger than in most reports (including our own \cite{noi3}); on the theory side, we have applied a simple correction that offers no guarantee of being equally appropriate for all transitions and all $x$. As we now discuss, an interesting crossover behavior  is more easily observable in the band offsets at the interface with the parent compounds.

\section{Band offsets across the phase transition}

Band offsets at interfaces are key quantities in the design and engineering of heterostructures. Ab initio theory has been predicting reliable offsets all along; a review of the theory concepts and of the  techniques in common use is in Ref.\cite{peressi}. Recently we predicted a staggered B-type offset at the (100) interface of $\beta$-phase Ga$_2$O$_3$ with low-$x$ (Ga$_{1-x}$In$_x$)$_2$O$_3$. Here we extend the work to a much larger range in the bixbyite structure. We simulate the (001)-like  interface with  In$_2$O$_3$/(Ga$_{1-x}$In$_x$)$_2$O$_3$ superlattices  epitaxially constrained to In$_2$O$_3$, containing 160 atoms in the primitive cell, and with explicit atomic substitutions.  We also calculate the same quantities relaxing all lattice parameters of the superlattice; this mimics a substrate that is locally compliant, i.e. deforms along with the film. For  $x$ inside a phase separation region,  the calculated values refer to the mixed phase, and not to the potentially   compositionally (or structurally) segregated one (see Sec.\ref{phsep}).

\begin{figure}[ht]
\centering
\includegraphics[width=9cm]{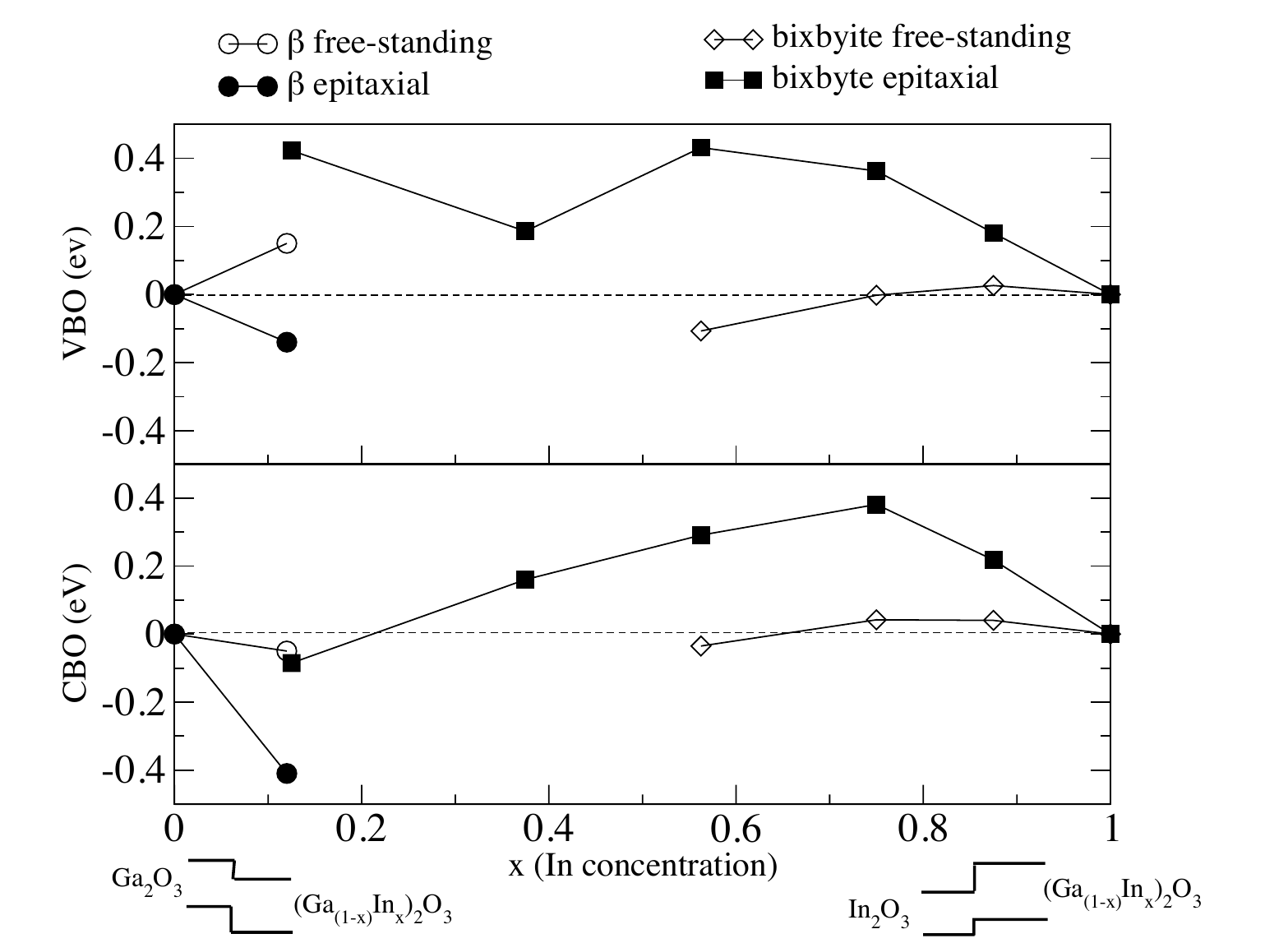}\qquad
\caption{\label{fig5} Valence (VBO, top) and conduction (CBO, bottom) interface band offsets between In$_2$O$_3$ and (Ga$_{1-x}$In$_x$)$_2$O$_3$ in the bixbyite phase, both epitaxially constrained on  In$_3$O$_3$, or with compliant substrate. The offset between Ga$_2$O$_3$ and low-$x$ (Ga$_{1-x}$In$_x$)$_2$O$_3$ at $x$$\simeq$0.1 in the $\beta$ phase for the same conditions  are also reported.}
\end{figure}

On the bixbyite side of the phase separation region, the offsets are again type-B (a relatively uncommon occurrence in itself), but most interestingly they are staggered in the opposite direction, i.e. both the conduction and valence  offsets encountered in going from the parent compound into the alloy are generally positive, whereas they were negative in the  low-$x$ limit (see the sketches in Fig.\ref{fig5}). This suggests interesting perpectives for interface offset tuning in this alloy system.  The offset values are also rather large, and hence interesting in terms of potential charge confinement. Recalling the phase separation region discussed above, the most promising $x$ is around 0.7 in the epitaxial bixbyite.
We note that the gap error of density functional theory is immaterial here, as any gap corrections will largely cancel out of the offsets themselves. We thus  purposely refer to offsets only,  starting from zero at $x$=0 and $x$=1. 

\section{Conclusions}	

Using  (mostly) density-functional ab initio  theoretical techniques, we have established that   (Ga$_{1-x}$In$_x$)$_2$O$_3$  will exist in the $\beta$ phase at low $x$ and in the bixbyite phase at high $x$. For the epitaxial bixbyite case, the compound will phase-separate above 15\%, and the  two phases will coexist up to about 55\%. While the free-standing bixbyite's coexistence with  the $\beta$ phase  is limited to about $x$=0.25, but bixbyite will itself separate into Ga-rich and In-rich regions for $x$ up to about 0.85. Thus the effective miscibility gap extends all the way from $x$=0.15 to $x$=0.85.  Importantly, both  miscibility gaps are practically independent of temperature, and survive up to the melting temperature. 
The behavior of the calculated volume, gap (in decent agreement with experiments for films), and band offsets also confirm the picture.  
Interestingly, we find that  the interface band offsets are largely type-B staggered and positive at large $x$, whereas they are staggered and negative in the low-$x$ limit.

\ack
Work  supported in part by MIUR-PRIN 2010 project {\it Oxide}, CAR of University of Cagliari, Fondazione Banco di Sardegna grants, CINECA computing grants.
MBM acknowledges  the financial support of her PhD scholarship by Sardinia Regional Government under P.O.R. Sardegna F.S.E. Operational Programme of the Autonomous Region of Sardinia, European Social Fund 2007-2013 -
Axis IV Human Resources, Objective l.3, Line of Activity l.3.1.


\section*{References}
\begin{thebibliography}{9}
\bibitem{noi1} Maccioni M B, Ricci F, and Fiorentini V 2015 {\textit Appl. Phys. Express} {\bf 8} 021102.
\bibitem{noi2} Maccioni M B, Ricci F, and Fiorentini V 2014 \textit{J. Phys. Conf. Ser.} {\bf 566} 012016.
\bibitem{vasp} Kresse G and Furthmuller J 1996 \textit{Phys. Rev. B}
\textbf{54} 11169.
\bibitem{sesqui1} 
Marsella L and Fiorentini V 2004 {\it Phys. Rev. B} {\bf 69} 172103.
\bibitem{sesqui2}
Delugas P,  Fiorentini V, and Filippetti A 2009 {\it 
 Phys. Rev. B} {\bf 80} 104301. 
\bibitem{baldini} 
 Baldini M, Gogova D,  Irmscher K, Schmidbauer M,  Wagner G, and Fornari R 2014 {\it 
Cryst. Res. Technol.} {\bf 49}, 552.
\bibitem{zhang} Zhang F, Saito K, Tanaka T, Nishio M, Guo Q 2014 \textit{Solid State Comm.} \textbf{186} 28.
\bibitem{galliaindia}
Edwards D D, Mason T O, Goutenoire F and Poeppelmeier K R 1997 {\it Appl. Phys. Lett.} {\bf 70} 1706.
\bibitem{noi3}
Ricci F, Boschi F, Baraldi A, Filippetti A, Higashiwaki M, Kuramata A, Fiorentini V, and Fornari R 2015, submitted for publication.
\bibitem{inogap}
King P D C, Veal  T D, Fuchs F, Wang C Y, Payne D J, Bourlange A, Zhang H, Bell G R, Cimalla V, Ambacher O, Egdell R G, Bechstedt F, and McConville C F,
2009 \PR B {\bf 79} 205211.
\bibitem{peressi}
Peressi M,  Binggeli N, and Baldereschi A, 1998 \JPD {\bf 31} 1273.

\end{thebibliography}
\end{document}